\documentclass[aip, amsmath, amssymb, reprint, superscriptaddress]{revtex4-1}
\usepackage{lipsum} 
\usepackage{graphicx}
\usepackage{dcolumn}
\usepackage{color}
\usepackage[normalem]{ulem} %strike out tool using \sout{...}
\usepackage{amsmath,amsfonts,mathtools}
\begin{document}
\title{Analysis of a single-mode waveguide at sub-terahertz frequencies as a communication channel}
\author{Marc Westig}
\email{mpwestig@gmail.com}
\affiliation{Kavli Institute of NanoScience, Delft University of Technology, 
Lorentzweg 1, 2628 CJ Delft, The Netherlands}
\affiliation{Fachbereich Physik, Freie Universit{\"a}t Berlin, Arnimallee 14, 14195 Berlin, Germany}
\author{Holger Thierschmann}
\affiliation{Kavli Institute of NanoScience, Delft University of Technology, 
Lorentzweg 1, 2628 CJ Delft, The Netherlands}
\author{Allard Katan}
\affiliation{Kavli Institute of NanoScience, Delft University of Technology, 
Lorentzweg 1, 2628 CJ Delft, The Netherlands}
\author{Matvey Finkel}
\affiliation{Kavli Institute of NanoScience, Delft University of Technology, 
Lorentzweg 1, 2628 CJ Delft, The Netherlands}
\author{Teun M. Klapwijk}
\affiliation{Kavli Institute of NanoScience, Delft University of Technology, 
Lorentzweg 1, 2628 CJ Delft, The Netherlands}
\begin{abstract}
We study experimentally the transmission of an electro-magnetic waveguide in 
the frequency range from 160 to 300 GHz.
Photo-mixing is used to excite and detect the fundamental 
$\mathrm{TE}_{10}$ mode in a rectangular waveguide with two 
orders-of-magnitude lower impedance. The  large impedance mismatch 
leads to a strong frequency dependence of the transmission, which we 
measure with a high-dynamic range of up to 80 dB, and with high 
frequency-resolution. The modified transmission function is directly 
related to the information rate of the waveguide, which we estimate 
to be about 1 bit per photon. We suggest that the results are applicable 
to a Josephson junction employed as a single-photon source 
and coupled to a superconducting waveguide 
to achieve a simple on-demand 
narrow-bandwidth free-space number-state channel. 
\end{abstract}
\maketitle 
\section{\label{sec:01}Introduction}
An important problem in communication technology is the 
transmission of  a coded message, from an initial point \emph{via} a channel to 
a final point, with minimal error in receiving and decoding the message. 
Electro-magnetic transmission lines, like an optical fibre or a 
waveguide, are attractive as a communication channel, because compared to 
free space, they minimize the radiative information loss.  
The effectiveness of a channel, in terms of the information rate, is 
quantified by the channel capacity, measured in bits 
per second or in bits per photon.\cite{caves1994, giovannetti2004,yuen1993}
The most effective way of using a waveguide is by implementing a 
"one quantum - one bit - one mode" strategy as introduced by \citet{caves1994}. 
Here, we analyze experimental results on the transmission of a waveguide 
at sub-terahertz (THz) frequencies in the context of this information 
communication-effectiveness. 

Three-dimensional (3D) waveguides and cavities have recently entered the field of 
quantum-information technology in the form of superconducting cavities, which 
show at the frequencies around 10 GHz the lowest (dissipative) loss 
to date\cite{Paik2011, Reagor2013} and, therefore, support a superior 
qubit operation. The reason for the low loss is the minimal relative energy 
stored in surface defects in 3D cavities,\cite{Turneaure1968} compared to 
other technologies like planar (2D) resonators or conductors.\cite{Barends2010} 
For a 3D cavity, the relative energy loss in surface defects scales inversely 
proportional to its size,\cite{Reagor2013} which is also dependent on the 
propagating mode, because of the mode-dependent current distribution at 
the cavity walls. This type of technology has been developed 
many years ago in the 
field of astronomical detectors\cite{Chattopadhyay2004} for 
applications at significantly higher frequencies. Developed for 
the astronomically important range of hundreds of GHz 
and for other sophisticated 3D sub-mm waveguide 
circuits,\cite{kazemi2015} the technology appears 
suitable for use in quantum networks as well.

In previous work, isotropic dispersion relations have been included
in the description of the communication  channel. In this case the 
signal group velocity along the channel at frequency $f$ scales with 
all spatial dimensions along the channel, as noted by \citet{caves1994} 
and, hence, the information rate is not modified by this isotropic dispersion. 
In the present work we identify that the frequency-dependent
transmission of a single-mode rectangular waveguide carries 
directly over to the channel capacity. We quantify this 
by analytic expressions and numerical modeling. It shows that it is 
caused by the high-impedance source and detector, coupled to the 
two orders-of-magnitude lower impedance of the 
fundamental $\mathrm{TE}_{10}$ mode of the waveguide 
(using a waveguide of length $l = 39.6~(79.2)$~mm, \emph{i.e.} 
which is 33~(66)-times longer than the center wavelength). 
As a result, for a single frequency-multiplexed coherent-state 
channel, the information rate varies up to a factor of about 2 over the 
transmission bandwidth of 210 to 300 GHz. 

The paper is organized as follows. Section~\ref{sec:02}
describes our experimental method and setup. In Sec.~\ref{sec:03A} and 
\ref{sec:03B} the electro-magnetic modeling results of the diagonal 
horn antennas and the waveguides of our setup are provided. 
In a second step we connect the modeling results to the optical properties of the 
used photo-mixers (Sec.~\ref{sec:03C}), 
i.e.~their characteristic wave impedance, 
using the ABCD-matrix formalism to obtain 
a complete model. The measured response is presented and 
analyzed in Sec.~\ref{sec:04}. In Sec.~\ref{sec:05} we 
summarize theoretical expressions for the channel capacity, in 
which the effect of the frequency-dependent transmission 
through our diagonal-horn antenna/waveguide assembly
on the information rate becomes clear. It includes essential details of the 
capacity for two types of communication channels, a coherent-state 
channel and a number-state channel. Section~\ref{sec:06} 
discusses experimentally realized Josephson circuits which together 
with superconducting waveguides are well suited for  
implementing in free-space the concepts 
discussed theoretically in Sec.~\ref{sec:05}. We conclude our work in 
Sec.~\ref{sec:07}.
\section{\label{sec:02}Experimental setup}
\begin{figure}[htb!]
\centering
\includegraphics[width=\columnwidth]{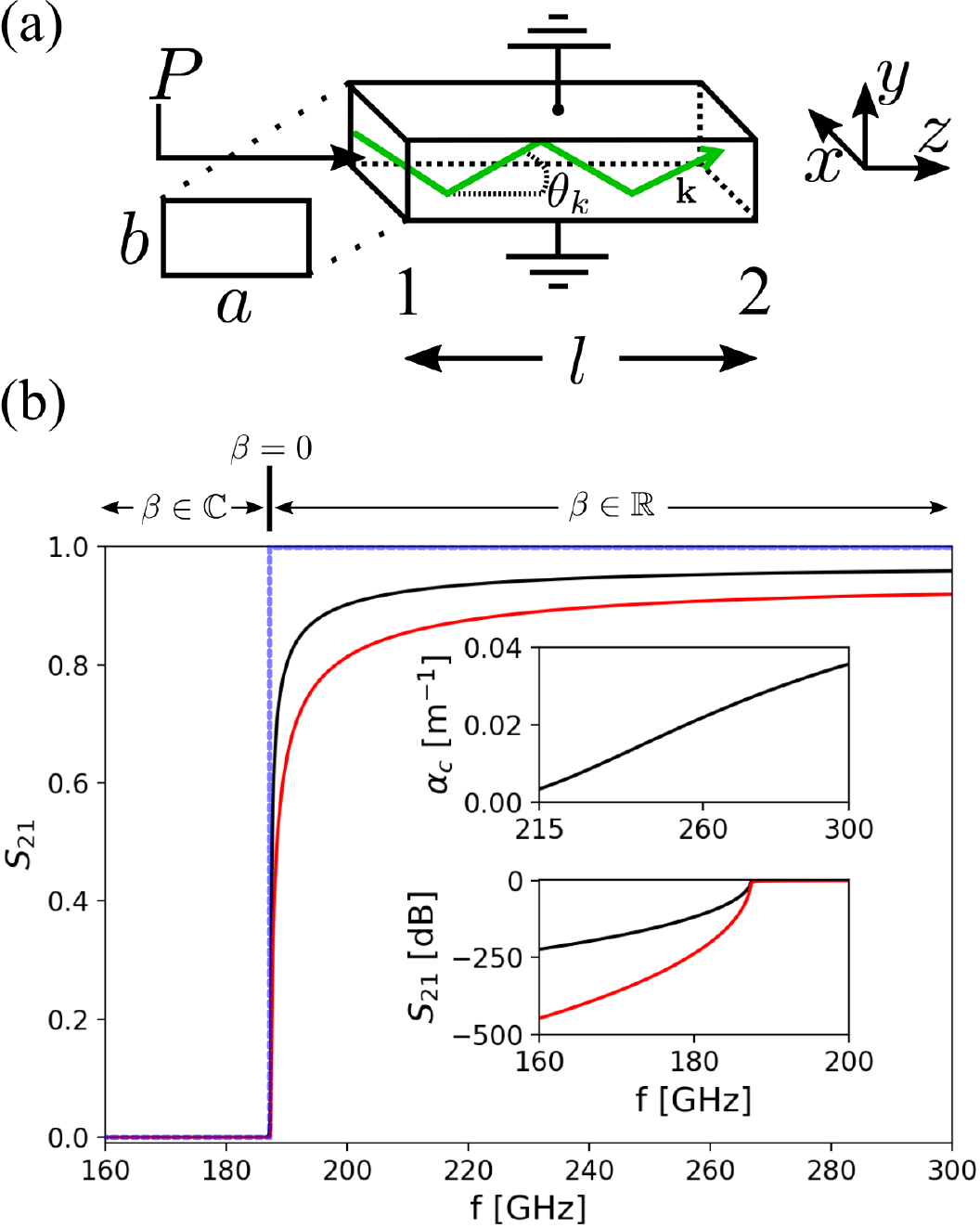}
\caption{\label{fig01}(a) Sketch of an ideal input/output 
impedance matched waveguide, assuming $a = 800~\mu$m 
and $b = a/2$. The signal power $P$ 
is injected at the input $1$ and reaches the output $2$.
If only the fundamental mode is excited, 
the polar angle $\theta_k$ quantifies via the relation
$c_{0}\cos(\theta_k)$ the longitudinal group velocity in 
the waveguide. 
(b) Calculated transmission from port 1 to port 2 for two 
different lengths $l = 39.6$~mm (black) and $l = 79.2$~mm (red) 
using the standard analytical theory presented in \citet{bayer1980}. 
The blue line is for an idealized response without loss as 
expected for a superconducting waveguide, as long as the 
photon energy is below the value of the superconducting 
energy gap. For frequencies lower than the waveguide 
cut-off frequency $f < f_{min} = c_{0}/(2a)$, the propagation 
constant for the fundamental mode, $\beta$, is complex-valued, 
meaning that only evanescent waves exist in the waveguide. 
Here, $k_{z} = 2\pi/\lambda_g$
is the wavevector with $\lambda_g$ being the guide 
wavelength. For $f_{min}<f<2f_{min}$, the propagation 
constant is real-valued and a traveling wave is launched in 
the waveguide characterized by the fundamental 
$\mathrm{TE_{10}}$ mode.
The upper inset in (b) shows the calculated conductor 
loss $\alpha_{c}$ for the material CuTe, used in the 
experiment and also for the calculations in this figure 
(red, black) as a function of frequency. The lower inset shows 
a magnified view of the evanescent-wave transmission range on a 
logarithmic scale.}
\end{figure}

\begin{figure*}[tb]
\centering
\includegraphics[width=\textwidth]{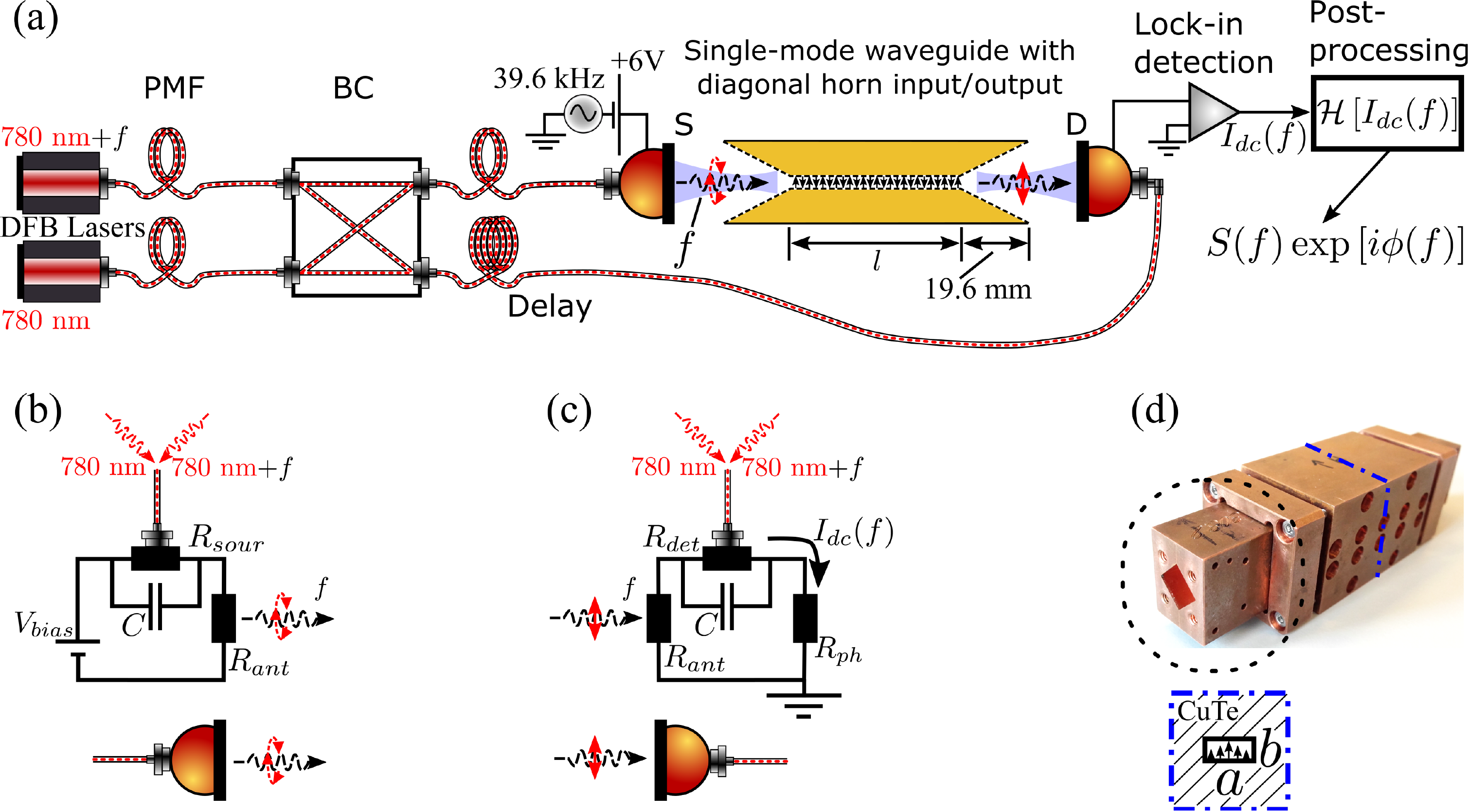}
\caption{\label{fig02}(a) Experimental setup. PMF = polarization 
maintaining fibres (Thorlabs GmbH), BC = beam combiner and 
S/D = source/detector (TOPTICA Photonics AG).\cite{toptica} The light blue 
profile indicates the Gaussian beam shape of the source and detector which 
is coupled to the diagonal-horn aperture. The dashed red circle at the input 
indicates the circular polarization of the source. The vertical red 
arrow at the output shows the linear polarization of the radiated field 
received by the detector. The polarization change occurs due to 
the electro-magnetic field distribution in the waveguide, sketched 
in (d). (b) and (c) indicate the electrical-circuit model of the 
source and detector. (d) Fabricated CuTe waveguide with mounted 
diagonal-horn antennas of the same material. The blue dash-dotted 
rectangle indicates a cut through the waveguide, sketching the hollow 
waveguide channel with length $a$ and width $b = a/2$ 
and the linearly polarized electrical field lines, parallel to the $b$-side.}
\end{figure*}

In Fig. \ref{fig01} we show a sketch of a standard open 
rectangular waveguide and the standard prediction for the 
transmission using the analytical theory presented in \citet{bayer1980}. 
Our aim is to evaluate experimentally this prediction for the 
sub-THz frequency range. The shown predictions are for 
a lossless and impedance-matched waveguide, 
the step-function, and for two impedance-matched waveguides 
of different lengths including the loss in the material. A longer 
waveguide naturally has a lower transmissivity because of the 
loss of signal occurring over the length of the waveguide. An 
experimental test of this prediction is currently possible because 
of the availability of high quality waveguides and continuously 
tunable sources.
The full experimental setup is shown in Fig.~\ref{fig02}(a). 
The central element is the waveguide, like schematically shown 
in Fig.~\ref{fig01}(a), and represented in detail in 
Fig.~\ref{fig02}(d). A free space sub-THz electro-magnetic 
wave enters a diagonal horn, travels through the waveguide, and is 
radiated out again from a diagonal horn. The horns and waveguide 
are made of the same material, CuTe, using present-day 
computer-controlled machining technology. The frequency-tunable 
electro-magnetic signal, in the appropriate frequency range of 160 to 
300 GHz, is generated and detected by superimposing the outputs of 
two 780~nm distributed feedback (DFB) lasers in a beam combiner (BC) 
and illuminating two GaAs photo-mixers connected at the output of the 
beam combiner via polarization maintaining fibers (PMF), with one 
photo-mixer acting as coherent THz source (S) and the second one 
as coherent THz detector (D).\cite{toptica} The incident laser power 
on each photo-mixer is approximately 30~mW. The desired frequency 
of the THz electro-magnetic signal is set by adjusting the 
difference frequency between the two DFB lasers. Optimal coupling 
between all optical elements is achieved by arranging the setup 
according to the distances summarized in Sec.~\ref{sec:03C}.

%The source (S) and detector (D) are two identical GaAs photo-mixers \cite{toptica}. 
%Both are driven by the same set of DFB-lasers (780 nm wavelength), using 
%polarization maintaining optical fibers (PMF) and a beam combiner (BC). 
Each photo-mixer consists of a metallic two-electrode log-spiral circuit, 
patterned on a GaAs chip. The non-patterned side of the GaAs chip is 
glued on a silicon lens, employed for Gaussian beam formation. The laser 
spot from the optical fiber is focussed on the feedpoint of the log-spiral circuit. 
%The combination of the log-spiral circuit and the silicon lens functions either as 
%an efficient THz emitter or detector, dependent on the way it is operated. 
The relevant equivalent circuits are drawn in  Figs.~\ref{fig02}(b) and (c). The source 
photo-mixer is biased by a 39.6~kHz modulated on/off voltage of 
$V_{dc} = 6\mathrm{V}$,  facilitating a lock-in detection of the measured signals. 
A detailed overview of the GaAs mixer-technology is provided in the reviews 
by \citet{preu2011}, \citet{brown2003} and \citet{saeedkia2013}. 

We analyze the emitted THz field from the diagonal horn 
as follows. The THz electric field component received by the detector 
leads to an ac-voltage drop across an interdigitated capacitor of the detector 
with a frequency equal to the laser detuning $f$. Together 
with the laser-induced impedance modulation with the same 
frequency, but in general with a different phase, 
a coherent dc-photocurrent, $I_{dc}(f)$, flows in the positive 
or negative direction (dependent on the phase) 
across the feedpoint of the log-spiral circuit. 
The dc-photocurrent induces a voltage drop across the detection 
impedance $R_{ph}$ (Fig.~\ref{fig02}(c)) which is amplified 
while  preserving its sign in a trans-impedance amplifier 
(type PDA-S) with a bandwidth of 0-1~MHz. Each data-point 
is then integrated over 500~ms. This detection scheme 
resembles a homodyne detector at THz frequencies with a 
high-dynamic range up to 80~dB \cite{toptica} like described 
by \citet{roggenbuck2010}. A benefit of this scheme is that 
it measures the transmitted amplitude rather than only the transmitted 
power. In this work we use only the instantaneous amplitude 
(envelope function $\propto \sqrt{P}$) of the Hilbert transformation 
of $I_{dc}(f)$, Eq.~(\ref{eq:01}), 
for comparison with our theoretical model:
\begin{equation}
\label{eq:01}
\mathcal{I}_{dc}(f) = I_{dc}(f) + i\mathcal{H}\left[I_{dc}(f)\right] 
= S(f)\exp\left[i\phi(f)\right]~.
\end{equation}
Here, $\mathcal{H}(\cdots)$ is the Hilbert transformation,\cite{vogt2017} 
$\phi(f)$ the instantaneous phase of the signal and $S(f)$ 
is the instantaneous amplitude which we show in 
Figs.~\ref{fig05} and \ref{fig06} as experimental data.

Since the interplay between 
generation and detection of the 
sub-THz signal plays a crucial role in 
our work, we address both in a bit more 
detail. At the feedpoint of the log-spiral 
antenna, the two electrodes are coupled to a few micrometer size 
metal-semiconductor-metal (MSM) interdigitated capacitor, having 
a square geometry. It functions as a photoconductive switch, since 
the GaAs chip becomes slightly conducting in the region of the 
interdigitated capacitor when charge carriers are created by the laser 
light, followed by recombination with the emission of phonons, 
characterized by a measured time scale of $\tau_{c} = 500$~fs. 
The interdigitated capacitor fingers are about 1~$\mu$m apart. 
The dc-resistance of the photoconductive switch is typically of the 
order of $R_{sour, det} = 10 \text{-} 30~\mathrm{k}\Omega$ \cite{toptica} 
when the switch is closed, dependent on the charge carrier density, 
tuned by the laser power. For the electro-magnetic 
description of the photoconductive switch of the source 
and the detector, the characteristic wave 
impedance is much higher and amounts to 
$R_{c}^{S,D} \sim 140~\mathrm{k}\Omega$.  It is estimated 
over the generated power in the frequency range used and the 
resulting current flow upon detection.\cite{privatecomm_TOPTICA} 
We  assume that the characteristic impedance contains 
no reactive components, 
which appears adequate to describe  our experimental results.

The initially uncorrelated laser fields are coupled by the nonlinear, 
i.e.~quadratic, dependence of the charge carrier creation in the 
GaAs (causing the frequency mixing). Detuning the DFB lasers by the 
frequency $f$, the impedance of the MSM interdigitated 
capacitor is modulated by the same frequency. 
In other words, the switch is opened and closed on a 
timescale $\sim (2\pi f)^{-1} \gg (2\pi f_c)^{-1} $, which in our 
frequency range is much slower than the characteristic 
timescale $(2\pi f_c)^{-1} = \tau_{c}$ of the recombining 
carriers in the GaAs. Furthermore, the antenna 
impedance of the log-spiral circuit, 
$R_{ant} \approx 72~\Omega$,\cite{McIntosh1995} 
together with the photo-mixer capacitance $C$, 
defines an $RC$-time, which in our used frequency range is 
much shorter than the time period of the generated 
waves due to the frequency mixing, $RC \ll (2\pi f)^{-1}$.

Hence, together with the application of the dc-bias voltage, 
a THz current with frequency $f$ oscillates in the MSM 
interdigitated capacitor and excites the log-spiral circuit. 
The log-spiral circuit has a counter-clockwise 
direction of turn in the propagation direction of the beam.
Due to this geometry, it emits a left-circular polarized THz 
field through the silicon lens, indicated by the dashed 
red circle in Figs.~\ref{fig02}(a) and (b).
Sending the left-circular polarized THz field through 
the diagonal-horn antennas and the rectangular 
waveguide, shown in Figs.~\ref{fig02}(a) and (d), 
the output field is linear polarized (red linear arrow 
at the output of the waveguide in Fig.~\ref{fig02}(a)). 
The reason for this is the confined parallel-plate like geometry 
of the rectangular waveguide, shown in Figs.~\ref{fig02}(a) and (d). 
In Fig.~\ref{fig02}(d) the cut through the full-height rectangular 
waveguide indicates the hollow region with area $2b^2$ 
machined in the CuTe material. For the fundamental 
$\mathrm{TE}_{10}$ mode, this 
geometry only supports field lines perpendicular to the 
a-side and parallel to the b-side, hence, an electro-magnetic 
field with a linear polarization is characteristic for this 
waveguide mode. Due to the conversion from a circular-polarized 
to a linear polarized mode in the waveguide, the output power 
is lower by a factor 2. It is further decreased by additional, 
albeit small, waveguide losses when considering that 
the traveling distance through the waveguides with lengths 
$l = 39.6$~mm and $l = 79.2$~mm corresponds to 
many wavelengths ($\lambda \sim 1.2$~mm) in our frequency range.

The detector is operated without applying a bias voltage to it 
and it is pumped by the same two DFB lasers through the beam 
combiner, like the source. This modulates again the 
impedance of the MSM interdigitated capacitor at the frequency 
$f$, but this time in the detector. The impinging THz electro-magnetic 
field on the detector is received by its log-spiral circuit and is 
characterized by an amplitude and phase which we obtain 
by post-processing using a Hilbert transformation of the 
real-valued detector photocurrent 
$I_{dc}(f)$, Eq.~(\ref{eq:01}).
\section{\label{sec:03}Electro-magnetic modeling and quasi-optical properties}
The measured data will be compared with model-data in Fig.~\ref{fig06}. 
Therefore, we describe first the electro-magnetic model which captures 
the full system. It is built from a separate analysis of the diagonal 
horn antennas (Sec.~\ref{sec:03A}) and the rectangular 
waveguide (Sec.~\ref{sec:03B}). These separate parts are 
then combined to a full model of the assembled device, including the 
coupling to the source and the detector (Sec.~\ref{sec:03C}). 
We use the standard ABCD-matrix 
formalism to obtain for each frequency the scattering parameter which 
subsequently can be compared to the measured transmission, 
cf.~Fig.~\ref{fig06}. The details of this approach are given below.
\subsection{\label{sec:03A}Diagonal-horn antenna}
A three-dimensional finite element simulation of the 
diagonal-horn antennas has been conducted based 
on the CST software.\cite{cst} The optimization 
of the antenna performance is achieved 
by varying the inner-conductor shape 
and length in the simulation. To obtain a compact 
device and to minimize the dissipative loss, a short 
antenna is favored, provided it has still a low enough 
return loss and reasonable quasi-optical beam properties. 
In the CST model the antenna design is built upon 
mesh cells, with the metallic boundary conditions 
imposed to define the electro-magnetic properties. 
The feedpoint (cf.~Fig.~\ref{fig04}, upper panel) has been implemented with 
an ideal impedance-matched waveguide port to minimize 
parasitic reflections. The waveguide port is also used as 
the excitation source for the simulation. The output (or input, 
when the antenna is employed as receiving device) 
of the diagonal-horn antenna is terminated by the 
free space impedance $Z_{\infty} = \mu_0 c_{0}$, 
with $\mu_0$ being the vacuum permeability 
and $c_{0}$ the velocity of light in vacuum. 
A Vivaldi shape with a smooth transition to
straight edges at the output, cf.~Fig.~\ref{fig04} upper panel, 
yields a minimum input reflection $S_{11}(f)$ 
over the broad range of frequencies used in the experiment.
\begin{figure}[tb]
\centering
\includegraphics[width=\columnwidth]{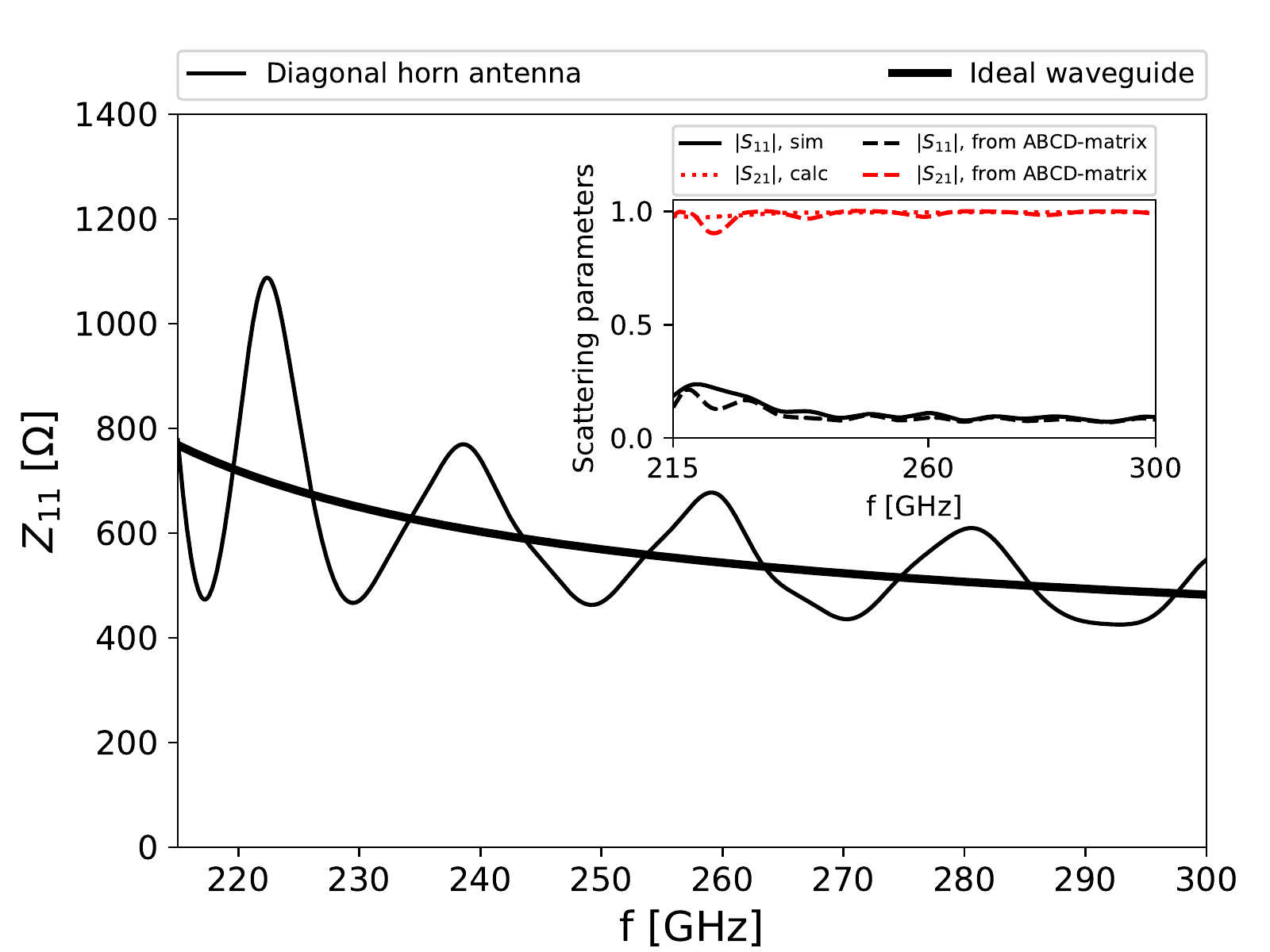}
\caption{\label{fig03}Numerical 3D finite-element 
simulation results of the diagonal-horn antenna 
with rectangular waveguide-feed
(Figs.~\ref{fig02}(a) and (d) and Fig.~\ref{fig04}), 
represented in terms of the input 
impedance $Z_{11}(f)$. For comparison, 
$Z_{11}(f)$ is shown also for an ideal impedance-matched 
rectangular full-height waveguide like 
sketched in Fig.~\ref{fig01}(a). In the 
simulation, the diagonal-horn antenna is terminated in 
the free space impedance $Z_{\infty} = \mu_{0} c_{0}$ 
and the waveguide is terminated by its 
frequency-dependent input impedance, 
representing the ideal impedance-match situation. 
The waviness in the result for the diagonal-horn 
antenna compared to the ideal waveguide result, 
shows the expected imperfection for our system 
when matching a rectangular waveguide via our 
diagonal-horn antenna to free space. The inset 
shows the simulated input reflection scattering 
parameter $S_{11}$ and the resultant (calculated) 
scattering parameter $S_{21}$ from the antenna 
feedpoint to its output as black solid and red short-dashed lines. 
The corresponding scattering-parameter pair derived 
from the ABCD-matrix formalism is shown for 
comparison as long-dashed lines.}
\end{figure}
\begin{figure}[tbp]
\centering
\includegraphics[width=0.5\columnwidth]{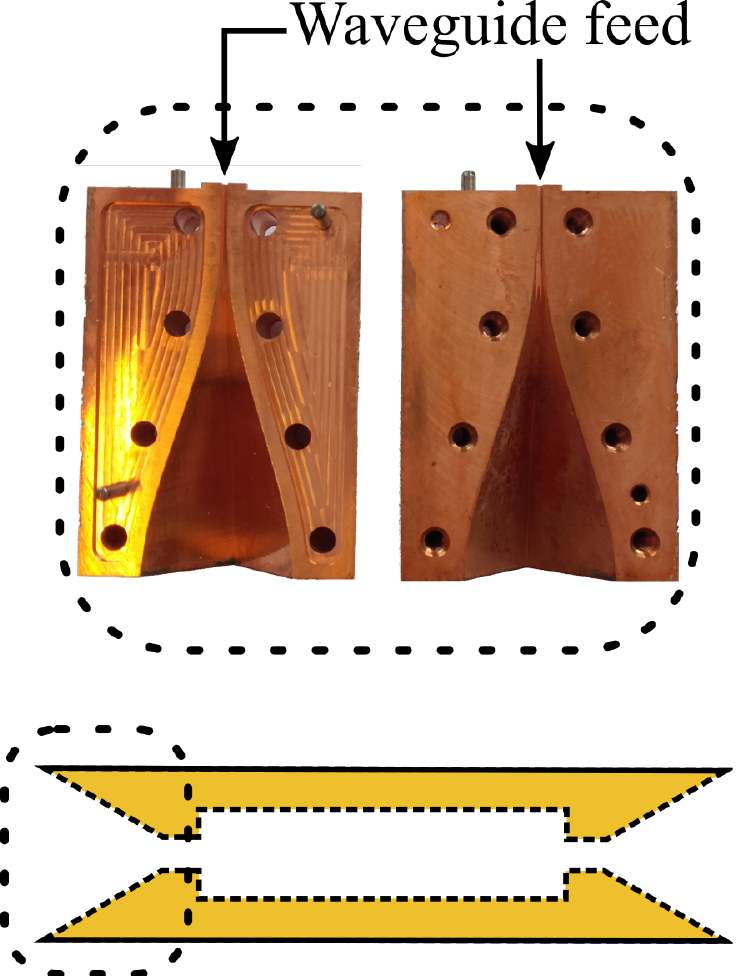}
\caption{\label{fig04}Adjustment of the ideal model to the experimental 
situation. Diagonal-horn antenna opened along the $E$-plane (top). 
The microwave surface currents are induced parallel to this plane, 
i.e.~perpendicular to the electric field. The bottom part 
of the figure sketches the cross-section through a diagonal 
horn antenna and waveguide assembly in which the waveguide 
channel is larger than the diagonal-horn antenna waveguide 
feedpoint. The interpretation of our measurements suggests such a 
configuration, which is in contrast to the ideal situation 
shown in Fig.~\ref{fig01}(a).}
\end{figure} 

As a final step, we have included the designed 
antenna geometry in a three-dimensional 
construction model prepared in AutoDesk Inventor.\cite{inventor}
\footnote{The construction model was used as blueprint 
for the micro-machining in the workshop of the 
I.~Physikalische Institut of the Universit{\"a}t zu K{\"o}ln.}
Through this we achieve a one-to-one correspondence between 
the anticipated optical properties in the simulated diagonal 
horn antennas and the optical properties of the fabricated 
diagonal-horn antennas. 

Our simulation results are summarized in Fig.~\ref{fig03}. 
In the main figure we show the input impedance $Z_{11}(f)$, 
the inset shows $S_{11}(f)$ as a black solid line. Both 
quantities are referred to the feedpoint. Note the waviness 
of $Z_{11}(f)$ of the diagonal-horn antenna, which indicates a 
less than perfect match to the free-space impedance. The 
waviness can be reduced at the cost of a longer antenna 
with a smoother transition from the feedpoint waveguide 
to its output. As an illustration, we compare in the same 
figure the input impedance of the diagonal-horn antenna 
and the one of an input-matched rectangular full-height 
waveguide with the same dimensions as the antenna 
feedpoint. Obviously they follow the same trend, but the 
impedance function of the waveguide is smooth. These 
simulations take into account the conductor loss $\alpha_c$ 
of the diagonal horn and waveguide material CuTe 
(inset of Fig.~\ref{fig01}). 

We include the expected dissipative loss in the simulation for the 
diagonal-horn antennas, although it is known to be rather 
low in this antenna type. The reason is the advantageous 
field distribution,\cite{johansson1992} which drives surface 
currents parallel to the mechanical cut shown in Fig.~\ref{fig04}. 
The calculated realized gain and the directivity 
in the antenna simulation is used to 
determine the antenna efficiency. From this a 
dissipative loss in the antenna of -0.53~dB in the center of 
the frequency band is found, small enough to approximate 
a reciprocal network. This approximation facilitates
subsequent modeling. The input-reflection
scattering parameter, $S_{11}(f)$, and the scattering parameter 
from the feedpoint to the antenna output, $S_{21}(f)$, can be 
expressed as,
\begin{equation}
\label{eq:02}
\begin{split}
S_{11}(f) &= \lvert{S_{11}(f)}\rvert\exp{\left[i\theta(f)\right]} \\
S_{21}(f) &= \sqrt{1-\lvert S_{11}(f)\rvert^2}\exp{\left[i\phi(f)\right]}~.
\end{split}
\end{equation}
In the equations, $\theta(f) = \arctan\lbrace\mathrm{Im}[S_{11}(f)]/
\mathrm{Re}[S_{11}(f)]\rbrace$ 
is the argument of the complex-valued function 
$S_{11}(f)$ and $\phi(f) = \theta(f)+\pi/2 \mp n\pi$.\cite{collin1992} 
Due to the reciprocal-network approximation, the 
Eqs.~(\ref{eq:02}) describe the antenna as an emitting as well 
as a receiving device, $S_{21} \approx S_{12}$. For the 
back-to-back configuration shown 
in Fig.~\ref{fig01}(a), $S_{12}$ is the scattering 
parameter from the input to the antenna feedpoint 
for the receiving antenna,  
which connects to the waveguide. This means in 
particular, deriving $S_{11}$ as a function of frequency 
is sufficient to determine the full set 
of scattering parameters for the diagonal-horn 
antenna.

Since the diagonal-horn antenna and waveguide device consists of 
three elements, a receiving diagonal-horn antenna, a rectangular 
full-height waveguide of length $l$ and an emitting and identical output 
diagonal-horn antenna, we combine in 
Sec.~\ref{sec:03C} the electro-magnetic 
properties of the three elements into a full model. A convenient 
way to achieve this is to exploit the ABCD-matrix formalism, which 
permits the combination of an arbitrary number of (two-port) microwave elements 
to calculate the full set of scattering parameters for 
the complete microwave elements assembly. The transformation 
between ABCD- and $S$-parameters for a two-port network 
with complex termination impedances is given by the matrix 
elements\cite{frickey1994}
\begin{equation}
\label{eq:03}
\begin{split}
A &= \frac{\left(Z_{01}^{*} + S_{11}Z_{01}\right)
\left(1-S_{22}\right) + S_{12}S_{21}Z_{01}}
{2S_{21}\sqrt{R_{01}R_{02}}} \\
B &=\frac{\left(Z_{01}^{*} + S_{11}Z_{01}\right)\left(Z_{02}^{*} 
+ S_{22}Z_{02}\right)-S_{12}S_{21}Z_{01}Z_{02}}
{2S_{21}\sqrt{R_{01}R_{02}}} \\
C &= \frac{\left(1-S_{11}\right)\left(1-S_{22}\right)-S_{12}S_{21}}
{2S_{21}\sqrt{R_{01}R_{02}}} \\
D &= \frac{\left(1-S_{11}\right)\left(Z_{02}^{*}+S_{22}Z_{02}\right)+S_{12}S_{21}Z_{02}}
{2S_{21}\sqrt{R_{01}R_{02}}}
\end{split}
\end{equation}
and
\begin{equation}
\label{eq:04}
\begin{split}
S_{11} &= \frac{AZ_{02} + B - CZ_{01}^{*}Z_{02} - DZ_{01}^{*}}
{AZ_{02}+B+CZ_{01}Z_{02}+DZ_{01}}\\
S_{12} &= \frac{2\left(AD-BC\right)\sqrt{R_{01}R_{02}}}
{AZ_{02}+B+CZ_{01}Z_{02}+DZ_{01}}\\
S_{21} &= \frac{2\sqrt{R_{01}R_{02}}}
{AZ_{02}+B+CZ_{01}Z_{02}+DZ_{01}}\\
S_{22} &= \frac{-AZ_{02}^{*}+B-CZ_{01}Z_{02}^{*}+DZ_{01}}
{AZ_{02}+B+CZ_{01}Z_{02}+DZ_{01}}~.
\end{split}
\end{equation}
In general, in the equations above every parameter is also a 
function of frequency which we include in our modeling. In addition,  
$Z_{01}$ and $Z_{02}$ are the complex termination impedances. 
The $*$-symbol indicates the 
complex conjugate and $R_{01}$ and $R_{02}$ are the 
real parts of the complex termination impedances.

By applying Eqs.~(\ref{eq:03}) we find the frequency dependent 
ABCD-parameters of the diagonal-horn antenna 
from the previously determined set of 
$S$-parameters. In this calculation, 
$Z_{02} = Z_{\infty}$ is the free space impedance 
terminating the diagonal-horn antenna and 
$Z_{01} = (\omega/c_{0}) Z_{\infty}/\beta$, 
with $\omega$ the angular frequency and 
$\beta$ the feedpoint waveguide 
complex propagation constant. We use 
first the nominal dimensions 
of the antenna's feedpoint waveguide 
to evaluate $\beta = \sqrt{(\omega/c_0)^2 - (\pi/a)^2}$, 
$a = 2b = 800~\mu$m.
By back-transformation to the $S$-parameters using 
Eqs.~(\ref{eq:04}), we evaluate the precision of this procedure in the
inset of Fig.~\ref{fig03} (long-dashed lines) with the previously 
simulated $S_{11}$- and calculated $S_{21}$-parameter 
(solid and short-dashed lines). 
In general we find a satisfactory agreement for all frequencies, 
the agreement is better for increasing frequency.  
At the lowest frequencies the deviation amounts 
to at most 8\%, accurate enough to permit a comparison to our 
experimental results.  

\subsection{\label{sec:03B} Rectangular waveguide} 
The standard ABCD-parameters for the rectangular 
waveguide are \cite{collin1992}
\begin{align}
\label{eq:05}
\begin{aligned}
A_{wg} &= \cosh{\left(\gamma l\right)} \\
C_{wg} &= Z_{01}^{-1}\sinh{\left(\gamma l\right)}
\end{aligned}
&&
\begin{aligned}
B_{wg} &= Z_{01} \sinh{\left(\gamma l\right)}\\
D_{wg} &= \cosh{\left(\gamma l\right)}~.
\end{aligned}
\end{align}
They take the same form as for the usual transverse-electro-magnetic 
(TEM) transmission line, like a coaxial cable. The only difference 
is the value of the complex propagation constant $\gamma$,   
due to the cut-off in the frequency spectrum of $\beta$ and 
its frequency dependence, 
$\gamma = \alpha_{c} + i\beta$, with $\alpha_{c}$ and $\beta$ as 
discussed above. 

\subsection{\label{sec:03C}Complete setup and quasi-optics}
Multiplying the ABCD-matrices of the single 
elements in the order as they appear in the setup shown in 
Fig.~\ref{fig01}(a), i.e.~receiving diagonal-horn antenna - rectangular 
waveguide of length $l$ - emitting diagonal-horn antenna, one obtains 
the full ABCD-matrix of the setup. Using the relations in 
Eqs.~(\ref{eq:04}) for the $S$-parameters, with 
$R_{c}^{S,D} = Z_{01,02} = 140~\mathrm{k}\Omega$ 
(Sec.~\ref{sec:02}) the real-valued 
source and detection characteristic wave 
impedances, one obtains the total $S$-parameter 
set of the experimental setup, which should be proportional to the 
transmission we measure in the experiment.

Finally, we introduce the experimentally 
relevant Gaussian, i.e. quasi-optical, beam 
properties\cite{goldsmith1998} of 
the source/detector and the diagonal-horn antenna, which is 
important in optimizing the experimental setup.    
First, the Gaussian beam profile of the source 
and detector are slightly asymmetric in their $xz$- and 
$yz$-planes, $z$ being the propagation direction 
of the field shown in Fig.~\ref{fig01}(a). This is quantified 
by the beam waist radii in these planes, 
$w_{xz} \sim 2$~mm and $w_{yz} \sim 2.2$~mm
which are located at distances $\sim 25$~mm 
and $\sim 32$~mm from the aperture. The optical 
properties of the diagonal-horn antenna, although having
slightly curved instead of linear slanted walls, is 
best approximated by the analytical theory 
by \citet{johansson1992}. From their theory, we obtain 
for our diagonal-horn antenna design a waist radius of 
$w_{xz} = w_{yz} = 1.9$~mm, a distance of $z_{A} = 14$~mm 
between the position of the beam waist and the aperture, as 
well as a Rayleigh length of $z_{c} = 9.6$~mm. 
Additionally, the beam curvature at the 
aperture is characterized by the radius $R_{A} = 20.6$~mm and 
the beam waist at the aperture is $w_{A} \approx 1.77w_{0}$. 
For maximum optical coupling we arranged the 
photo-mixers and the diagonal-horn antennas such that the 
respective beam waists were on 
top of each other. Optimizing the position of the 
photo-mixers with respect to the diagonal-horn antennas 
has been achieved by moving them with micro-meter precision 
in order to maximize the photocurrent and to compensate for 
their aforementioned beam asymmetry. 

\section{\label{sec:04}Experimental results and analysis}
\begin{figure}[tbp]
\centering
\includegraphics[width=1\columnwidth]{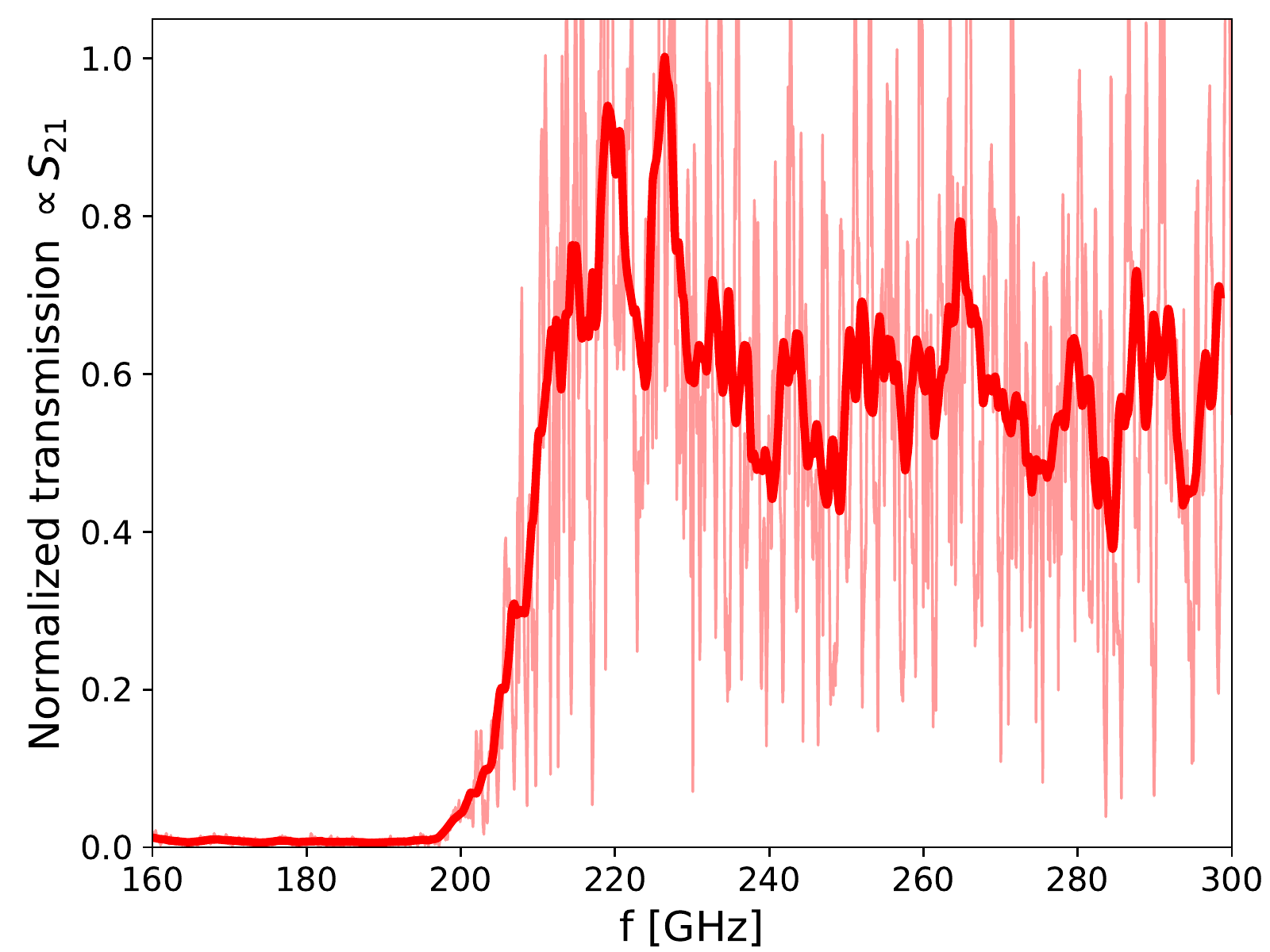}
\caption{\label{fig05}Raw data (thin transparent line) 
to averaged data (thick line) processing for the case 
of the long-waveguide, cf.~Fig.~\ref{fig06}(a). 
In this case the raw data comprises six traces which were acquired 
subsequently and then averaged whereas the thick line is the 
result of the moving average post-processing procedure as 
described in the text.}
\end{figure}

\begin{figure}[htbp]
\centering
\includegraphics[width=\columnwidth]{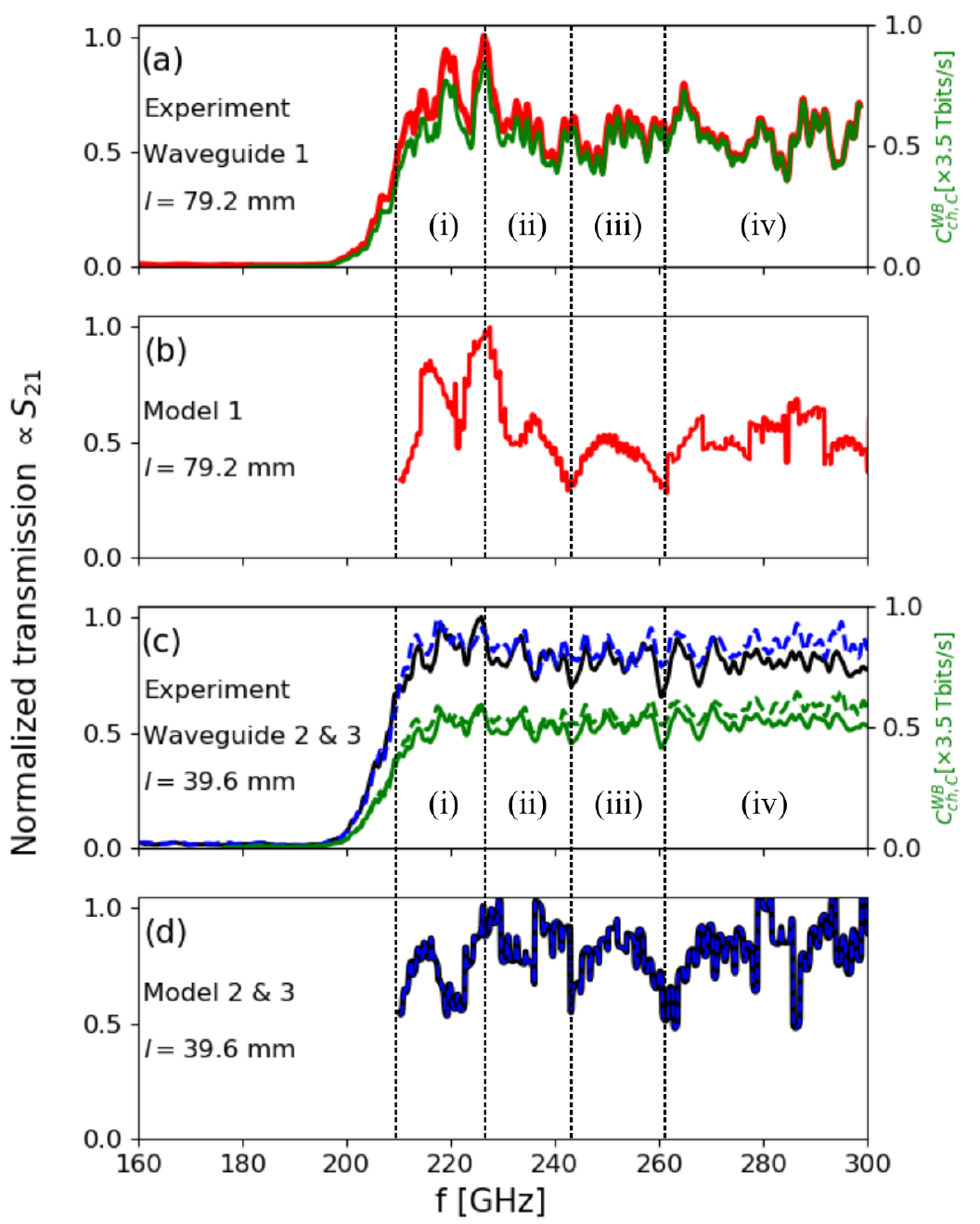}
\caption{\label{fig06}Waveguide 
transmission measurements and modeling, expressed 
as instantaneous amplitude which is proportional to 
the scattering parameter from port 1 to port 2, 
$S_{21}$, cf. Fig~\ref{fig01}(a).
The transmissions (left y-axis) 
are normalized and by this 
referred to the waveguide input, neglecting a 
finite but small dissipative loss of the transmitted 
electro-magnetic waves.
(a) and (c) show measurements of three 
different waveguides and two of these waveguides have 
the same length, shown in (c). (b) and (d) show modeling 
results for waveguides of two different lengths 
with $l = 39.6, 79.2$~mm used in the 
experiment. The wave transport through the waveguides 
with $l = 39.6$~mm and 79.2~mm is characterized 
by fundamentally different 
features over the measured frequencies 
and to explain this unambiguously, 
we label the data by sub-bands 
(i)-(iv), explained in more detail in Sec.~\ref{sec:04}. Note that the 
wave transport features are weaker for the short waveguides, (c), 
but that the most pronounced features are reproduced by the model 
in (d). The green traces in (a) and (c) summarize the calculated 
wideband coherent-state channel 
capacity [Eq.~(\ref{eq:08})] including the conductor loss. 
They are derived using the measured transmission, the power 
coupling factor determined by our model and an emitted power of 
$\sim 1~\mu\mathrm{W}$ at 
235~GHz, nearly constant over the other employed frequencies.}
\end{figure}

The full results from the measurements and the 
model analysis for three different waveguides are presented in 
Fig.~\ref{fig06}. Panel (a) (79.2 mm waveguide, experiment) 
and panel (b) (79.2 mm waveguide, model) are 
the transmission (proportional to $S_{21}$) 
as a function of frequency. Panel (c) and (d) show the results for the 
two short waveguides (39.6 mm), with panel (c) presenting the experiment and 
panel (d) the model. Obviously, all three waveguides 
become transmissive  around 210 GHz (panels (a) and (c)), 
the anticipated by design waveguide cut-off. 
In addition, it is also clear that the two short waveguides show a 
very similar pattern as a function of frequency, 
short-period oscillations as a function of frequency. 
Thirdly, we also observe that the twice longer waveguide, panel (a), shows 
more variation in the transmissivity as a function of frequency than 
the two shorter ones. These variations occur on a 
much longer frequency scale than the features observed in the 
two shorter waveguides. To proceed towards 
a quantitative evaluation, which in the end will provide the 
communication rate of the waveguide, we address first some 
aspects of the data-processing (cf.~also Fig.~\ref{fig05}). 
It is followed by evaluating the deviations between the 
design-values of the waveguides and the actual realizations. 
We then comment on the observed differences 
over the full band (210-300 GHz), which we arbitrarily divided into 
4 smaller sub-bands: 210-225 (i), 225-243(ii), 
243-261(iii) and 261-300 (iv) GHz. Finally, we infer from 
these quantitative results the communication rate. 

For the data presented in Fig.~\ref{fig06} 
the experimental and model results are both 
averaged, using a moving average post-processing 
procedure with an averaging length much smaller than the length 
of the data. The experimental data consist of six 
frequency-sweep datasets for the long waveguide and of two 
frequency-sweep datasets for each of the two short waveguides. 
The datasets for each waveguide were acquired subsequently 
and in the present analysis are averaged before the 
moving average post-processing procedure, 
cf.~Fig.~\ref{fig05}. In this way we smooth 
out fast fluctuations introduced by standing waves, 
caused by the large impedance mismatch 
between photo-mixers and diagonal-horn antennas. 
After this we normalize the data to enable a mutual comparison. 
The reason for this is that the setup shown in Fig.~\ref{fig01}(a) 
determines essentially a relative transmission in comparison 
to the situation when no field is transmitted through the 
waveguide. Unfortunately, a reference calibration procedure 
is generally difficult with this type of setup due to the sensitive 
optical alignment and the difficulty to measure 
absolute powers at these high frequencies. 
Nevertheless, the calculated $S$-parameters 
from the model permit an analysis 
of the absolute values of the signal loss of the transmitted field. 

We find that in order to bring consistency between the measurements 
and the modeling results, we have to assume a smaller rectangular 
waveguide feedpoint of the diagonal-horn antenna, than the waveguide 
connecting the two diagonal-horn antennas (as sketched 
in Fig.~\ref{fig04}). The most important 
experimental indication justifying this assumption is that for 
our three different waveguides, using the \emph{same} diagonal 
horn antennas, we measure consistently the same onset 
of transmission, of about $f_{min} = 200$~GHz. Since the 
waveguide fabrication is not reproducible with this precision, 
for instance a deviation within the machining precision of $\pm5$\% of 
the waveguide $a$-side would shift $f_{min}$ already by 
$\pm20$~GHz, this suggests that the rectangular waveguide 
feedpoint of the diagonal-horn antenna is the bottleneck for 
the transmission. Based on our measurements, it has to 
have a size of the order of 
$a = c_{0}/(2f_{min}) \sim 750~\mu$m. This corresponds 
to a deviation of $50~\mu$m from the nominal design of $800~\mu$m 
for the $a$-side introduced earlier and is consistent with the 
uncertainty expected from the waveguide machining. We assume, therefore, 
that the onset of transmission in all three waveguides is due to the size of the 
entry- and exit-orifice of the horn antennas.

Furthermore, in order to explain unambiguously the measured 
transmission spectrum we have to assume for the experimental 
waveguides slightly different dimensions than designed. For the long 
waveguide with length $l = 79.2$~mm a modified cross-section 
$a = 833.010~\mu$m. Similarly, for the shorter waveguides with 
lengths $l = 39.6$~mm a modified cross-section 
$a = 856.285~\mu$m. While choosing these new 
values for $a$ in the waveguide, we keep in the model 
the diagonal-horn antenna feedpoint waveguide fixed to the 
nominal value of $800~\mu$m for simplicity. In making the 
latter simplification, we accept a small underestimate of the 
input impedance of the diagonal-horn antenna, which turns out to be negligible 
for our analysis. Note that the decimal numbers
of the $a$-values indicate the exact input we used for the 
modeling. Although the decimal numbers specify just a small length 
variation relative to the wavelength, slightly changing them 
modifies also the modeling results. This result is expected for a 
microwave device which is much longer than the wavelength and 
has a large impedance mismatch. It causes a sizeable standing 
wave ratio which is strongly influenced even by such small relative length 
variation, modifying the electric field amplitude at the 
detector. Additionally, we assume that the waveguides are 
shorter by in total about $11~\mu$m. 
This amounts to 4\% of the height of the 
diagonal-horn waveguide flange which is mechanically 
squeezed during the mounting based on 
our experience, due to compression of the 
diagonal-horn antenna waveguide 
flange using the screws shown on the clamp in 
Fig.~\ref{fig01}(d) in the dashed area. This length modification has 
been included in our model through shortening the waveguide length 
$l$ in Eqs.~(\ref{eq:05}). With these adjustments the results shown 
in Fig.~\ref{fig06} can now be discussed including the comparison 
between experimental and modeled results. 

As mentioned before, we divide the results for the 
transmission in four sub-bands 
210-225 (i), 225-243(ii), 243-261(iii) and 261-300 (iv) GHz. 
This choice has no physical basis, but was suggested by the 
pattern observed in both the modeling and the experiment, in 
particular for the long waveguide. For this waveguide with length 
$l = 79.2$~mm, Figs.~\ref{fig06}(a) and (b), the transmission shows 
three maxima in the regions (i) and (ii) and is slightly flatter in 
regions (iii) and (iv), but still showing pronounced features. 
All observed features in the measurement in (a) can be 
qualitatively and mostly also quantitatively related to the pattern 
found by the model shown in panel (b). 
However, the peak which we observe in the measurement
between regions (ii) and (iii) is not found from the model. 
We assume that, given the large impedance mismatch between 
the photo-mixers and the diagonal-horn antenna, together with 
the overall length of the diagonal horn and waveguide device, 
much larger than the transmitted wavelength, that transmission 
features appear which are not exactly reproduced by our model. 
In this case the setup, but also the modeling, are very 
sensitive to small length variations, which would introduce 
similar features. We suggest that this is the most likely 
explanation for this particular transmission feature. 

For the shorter waveguides of length $l = 39.6$~mm, the 
overall transmission is much flatter and shows less pronounced 
maxima in the spectrum. Similar to the long waveguide, 
the regions (i) and (ii) in the model for the short waveguide, 
Fig.~\ref{fig06}(d), show three peaks. These 
can only be partly assigned in the 
measured transmission of waveguide 2, shown in 
Fig.~\ref{fig06}(c), and of waveguide 3 shown in the same 
panel. The transmission of the 
latter one (blue-dashed line) compared to 
waveguide 2 (black solid line) shows 
generally a smoother frequency response with less
pronounced features. In these waveguides, the transmission 
in the regions (ii) to (iii) is first slightly decreasing and 
then in region (iv) slightly increases again, separated 
from a small dip between region (iii) and (iv). At least 
this trend is reproduced by the model. This is clearly 
manifest for waveguide 2 for which also the dip is slightly deeper 
than for waveguide 3. In the latter case it is almost not distinguishable 
from the other ripples in the transmission. In region (i), the model 
predicts also a dip at around 220~GHz, which is 
not obtained in the measurement. Similar to the case for the long 
waveguide, we suspect here also the strong 
sensitivity to length variations of the waveguide,
leading to standing wave patterns in the setup.

We like to add that the short period 
oscillations in the short waveguides are reminiscent 
of universal conductance fluctuations (UCFs), studied 
earlier theoretically for light-scattering in 
waveguides.\cite{feng1988} In the latter work, 
the central outcome 
are correlation functions which quantify the coherence 
of the fluctuating intensity pattern of the light traveling 
through the waveguide, leaving out both inelastic scattering 
and absorption which have the effect of weakening or 
even destroying the coherence. The latter effects cannot 
be neglected in our case due to our normal conducting 
waveguides. Nevertheless, we built the auto-correlation function out 
of the raw data of the amplitude fluctuations measured 
in our short and long waveguides, and obtain a similar shape of the 
correlation function like derived in \citet{feng1988}, 
in our case, however, as a function of frequency lag, $\Delta f$, 
instead of wavevector lag.
We find that the amplitude auto-correlation functions for the short 
waveguides decay for $\Delta f = 492$~MHz to a value 
$1/e$, whereas for the long waveguide the same decay 
is already obtained for $\Delta f = 334$~MHz. The bare 
transmission between the two photo-mixers, i.e.~without a waveguide 
in-between, shows also a fluctuating transmission, most likely due 
to internal reflections in the Si lens and standing-waves, and needs to 
be disentangled. In this case $\Delta f = 550$~MHz takes the largest value. 
Interestingly, we find that the 
cross-correlation between the amplitude fluctuations of two independent 
measurements of the two different short waveguides yields a non-vanishing 
correlation, about half has large as the auto-correlation, obtaining 
$\Delta f = 900$~MHz and suggesting a strong coherence between them. 
Cross-correlating in the same way the amplitude fluctuations measured
in the long waveguide with the ones in the short waveguides, yields a much 
shorter coherence of $\Delta f = 200$~MHz and a much weaker correlation 
only about 17~\% of the auto-correlation. As mentioned in 
\citet{feng1988}, waveguides and microwaves are a versatile system to 
study UCFs, but concerning this aspect and the present status of our work, 
more work has to be done in the future.

Finally, we address the communication rate of these waveguides. 
We calculate, using the measured transmission, 
the expected communication rate through the waveguides, 
based on Eq.~(\ref{eq:08}) of the next section. 
The results for the different waveguides 
are shown as green traces in the panels along with the experimental 
results in Figs.~\ref{fig06}(a) and (c). To achieve these plots, the 
central challenge is to calculate the power $P$ in Eq.~(\ref{eq:08}) 
at the input of the waveguide. From our model we determine the 
electric-field coupling constant from the input to the output of the 
waveguide and the diagonal-horn antennas, i.e.~$S_{21}$,
and multiply it with the measured 
and then normalized transmission. Since the model 
accounts for dissipative losses, we obtain finally the absolute 
transmission. From this we calculate the power coupling and 
multiply it with the expected generated power at $235$~GHz, about 
$1~\mu$W,\cite{toptica, roggenbuck2010} taking into account 
the operation parameters of the THz source. Note that alternative direct 
and possibly absolute measurements of the power at these high frequencies 
are very difficult and often use the pyroelectric effect.\cite{sizov2018} 
This would introduce different uncertainities so that our procedure 
gives already a good enough estimate of the communication rate.

We finish the experimental section by comparing the possible 
communication rates we obtain, of the order of 1 bit per photon in 
the wideband coherent-state channel limit, 
with the limits derived by \citet{caves1994}. We obtain this result 
using the theoretical framework described in Sec.~\ref{sec:05} which 
also explains the assumptions used in our derivation.
From their theory, the maximum communication rate expected 
for a coherent-state channel is of the order of 2 bits per photon. 
They find that this maximum rate can only be achieved by 
experimentally realizing the "one quantum - one bit - one mode" 
strategy we have mentioned in the introduction. This would 
be the case if $P/h f \sim C \sim f$. In our 
system, however, $P/hf \gtrsim C \gg f$. From this result, 
the implication for the practical realizable communication rate 
can be drawn from the analysis in \citet{caves1994}. 
Our results imply that for our particular 
waveguide geometry, the injected power and used carrier frequencies, 
the rate at which quanta are transmitted down the channel is similar 
but still somewhat higher than the
channel capacity and the carrier frequency is the slowest quantity.
As a consequence, about one quantum per bit of information can be 
used, but far less than one bit per period (mode) is transmitted. This 
can be compensated by increasing the carrier frequency (at 
the same time reducing the waveguide dimensions to keep 
the single-mode character) and increasing the injected power 
$P$ with the goal to realize the ideal 
"one quantum - one bit - one mode" communication strategy.
\section{\label{sec:05}Channel capacities}
A rectangular waveguide, like shown in Fig.~\ref{fig01}(a), is often 
viewed as a component in a receiver system, which guides a 
signal to a detector. Alternatively, it can be analyzed in 
information theory as a channel to guide information, which can 
be decoded with a minimal amount of error. In 
classical physics, noise is introduced by signal 
loss in the channel or by coupling loss, such as  
when the signal is injected into the channel 
from an emitter. If quantum mechanics 
plays a role, for example in the case of a coherent receiver, 
noise due to vacuum fluctuations has to be taken into account. 
For a signal carrier frequency $f$ with bandwidth $\delta f$ a 
noise power of $\sim hf\delta f$ will contribute. 
This contribution will be significant for signals with high-frequency carriers 
and a wide bandwidth of the order of $\delta f \sim f$. Both 
contributions, classical and quantum, have the effect of reducing 
the signal-to-noise ratio (SNR) at the receiver.

The physics of information theory 
has been formulated by Shannon,\cite{shannon1948_01} 
creating the basis of  quantum information theory.\cite{bennett1998} 
The main interest is in the maximum rate at which information can 
be sent through a channel,  while being still decodable without error. 
A quantitative measure is the maximum channel capacity $C_{ch}$, 
expressed in bits/sec. It is derived from the information entropy in 
the Shannon-theory through the source and channel 
coding theorems.\cite{shannon1948_01} For the case of 
a lossless channel and a negligible thermal photon population, the 
classical channel capacity reads:\cite{shannon1948_01,caves1994}
\begin{equation}
\label{eq:06}
C_{ch,cl} = \delta f \log_{2}\left(1 + \frac{P}{N \delta f}\right)~,
\end{equation}
with $P$ the transmitted power through the channel and 
$N$ the noise power per Hz. In classical physics there is no 
lower limit on $N$, i.e.~the noise can be infinitesimal small. 
Consequently, the channel capacity would 
tend to infinity. This unphysical result has motivated a 
number of theoretical studies to determine the proper 
physical limit of the channel capacity, reviewed by 
\citet{caves1994} and by \citet{yuen1993}. For a 
quantum channel, cf.~the later 
Eq.~(\ref{eq:09}), \citet{yuen1993} 
derive in addition the limits of the average 
transmitted power $P$. Their work derives the maximum 
of the von Neumann information entropy, quantifying 
the maximum number of available input states to the 
channel, evidencing that it is classically an unbounded 
value but not quantum-mechanically. A heuristic 
method to impose a limit to $C_{ch,cl}$ is to state 
$N \sim hf$ so that,
\begin{equation}
\label{eq:07}
C_{ch,C} = \delta f \log_{2}\left(1 + \frac{P}{hf \delta f}\right)~.
\end{equation}

The subscript $C$ refers to a coherent state. 
The second part of the term between the brackets of Eq.~(\ref{eq:07})  
is then the mean photon 
occupation number, $\bar{n} = P/(hf\delta f)$,  and the power $P$ 
can be understood as a classical quantity. 
In this limit, Eq.~(\ref{eq:07}) gives already the correct 
channel capacity for a free-space single coherent state channel 
with narrow bandwidth, $\delta f \ll f$. This 
expression yields a maximum capacity of $\approx 0.586$ 
bits per photon. It can be determined by expressing Eq.~(\ref{eq:07}) as a 
normalized capacity, $C_{ch,C}\rightarrow C_{ch,C}/(\delta f \bar{n})$, 
followed by finding its maximum.\cite{caves1994}

In practice a wideband 
capacity is often relevant, for example, when many longitudinal 
modes are present within one transverse mode by 
frequency multiplexing. In this case, $\delta f \sim f$, and 
Eq.~(\ref{eq:07}) is for a coherent state no longer valid, since $\bar{n}\ll 1$. 
Including the frequency dependence of the photon energy, 
\citet{caves1994} derived the wideband limit of the single 
frequency-multiplexed coherent state channel, yielding the capacity
\begin{equation}
\label{eq:08}
C_{ch,C}^{WB} = \mathcal{A}\frac{1}{\ln(2)}\sqrt{\frac{2P}{h}}~,
\end{equation}
and, hence, $\mathcal{A}\times 2.0403$ bits per photon.

For a single number-state wideband channel, 
the channel capacity becomes
\begin{equation}
\label{eq:09}
C_{ch,n}^{WB} = \mathcal{A}\frac{\pi}{\ln(2)}\sqrt{\frac{2P}{3h}}~,
\end{equation}
and yields, therefore, $\mathcal{A}\times 3.7007$ bits per photon.
Equation~(\ref{eq:09}) is determined by application of the 
so-called Holevo bound,\cite{caves1994,holevo_in_other_works} 
which quantifies the maximal amount of information that 
can be obtained at the output of the channel using the 
concept of mutual information transfer. We like to add that the 
same Eq.~(\ref{eq:09}) has been derived in \citet{yuen1993} 
by relaxing the finiteness assumptions in the Holevo bound and 
considering instead an infinite-dimensional input/output alphabet 
for the communication. A detailed discussion about the history 
of these approaches is given in \citet{caves1994}.

The prefactor $\mathcal{A}$ has to be determined separately for different 
types of waveguides and field-patterns. In Eqs.~(\ref{eq:08}) and 
(\ref{eq:09}), $\mathcal{A} = 1$ for a free-space channel. For a 
full-height rectangular waveguide with dimensions $a=2b$, 
cf.~Figs.~\ref{fig01}(a) and \ref{fig02}(d), one gets $\mathcal{A} = \cos(\theta_{k})$. 
Moreover, $c_{0}\cos(\theta_{k})$ quantifies the reduction in 
longitudinal group velocity in the waveguide due to reflection 
at the waveguide walls (\citet{giovannetti2004} and Fig.~\ref{fig01}(a)). 
The angle of reflection is given by 
$\theta_{k} = \arccos\left(k_{z}/\lvert {\bf k}\rvert\right)$
for the fundamental $\mathrm{TE}_{10}$ mode, where 
${\bf k} = \left(\pi/2a,0,2\pi/\lambda_g\right)$, with $\lambda_g$ the 
wavelength in the waveguide. 
Note that the wavelength in the waveguide 
$\lambda_g > \lambda$ is always larger than the wavelength in free space, 
since $\lambda_g = 2\pi/\beta$ with $\beta = \sqrt{(\omega/c_0)^2 - (\pi/a)^2}$ 
the complex propagation constant. Consequently, $\theta_k \rightarrow \pi/2$ for long 
wavelengths approaching the waveguide cutoff 
$\lambda \sim 2a$, i.e.~$\omega/c_{0} \sim \pi/a$, where 
$\lambda_g$ diverges and $\beta \rightarrow 0$, resulting in 
a vanishing group velocity, $(d\beta/d\omega)^{-1} = c_0 \beta \lambda/2\pi$.  
In other words the channel capacity vanishes in this case since no 
information can be transmitted anymore. 

This overview of formulas for the channel capacities demonstrates that the 
value depends theoretically on the type of waveguide. A convenient way to 
study the predications of Eqs.~(\ref{eq:08}) and (\ref{eq:09}) experimentally 
is to realize a frequency-multiplexed channel, which splits the available 
bandwidth into small non-overlapping frequency bins, which together carry 
a total transmitted power $P$. 
Furthermore, dependent on the type of channel, 
one has to realize the optimal photon-occupation number for each bin in order to 
achieve the maximum capacities (\ref{eq:08}) and (\ref{eq:09}). 
Optimal photon-occupation number distributions 
for the number-state and coherent-state channel are summarized in \citet{caves1994}.
For a single-mode waveguide like studied by us, each frequency bin represents then 
an independent longitudinal channel with different $\theta_{k}$, 
within the same transversal mode, 
in this case the fundamental $\mathrm{TE}_{10}$ mode.

In the practical case of a lossy waveguide, like in our 
experiment, illustrated by the black and red transmission 
curves shown in Fig.~\ref{fig01}(b), not all 
photons of the electro-magnetic field supplied at the input of the 
waveguide will reach the output. For the capacity of a number-state 
channel this creates a difficult theoretical problem related 
to the exact form of the input information entropy. This problem is  
explained in detail by \citet{giovannetti2004} and recently in 
\citet{ernst2017} by considering a general quantum channel. 
This recent work studies as a proof-of-principle  
the regularization of the maximum pure-state 
input-output fidelity of a quantum channel. We expect, that for the 
capacity of a quasi-classical coherent state channel, 
Eq.~(\ref{eq:08}), reasonable conclusions are still possible 
even in the presence of loss. Therefore, we focus on this 
type of channel while conducting our experiment.

\section{\label{sec:06}Free space quantum optics 
employing Josephson photonics and waveguides}
This section addresses the conceptual framework of realizing 
free space quantum optics using recent achievements in Josephson 
photonics.\cite{rolland2019, gramich2013, dambach2015, 
grimm2019, leppakangas2015, leppakangas2016, westig2017, 
leppakangas2013, leppakangas2014, armour2015} 
In particular, the experimental works 
reported by \citet{rolland2019} and \citet{grimm2019} implement an 
only battery-powered Josephson junction which is strongly coupled to a 
single mode, realizing a single photon source. The experiment 
in \citet{westig2017} reports two-mode amplitude 
squeezing below the classical limit using also an only 
battery-powered Josephson junction, but this time weaker 
coupled to two modes of different frequency. It is important 
mentioning that \citet{grimm2019} realized a NbN based 
Josephson junction that could in principle generate 
radiation in the frequency window employed in our work, 
extending the frequency range compared to only 
aluminum-based circuits.

For a given frequency and an adequate 
choice of the waveguide and diagonal-horn antenna 
dimensions, we like to recall that the waveguide 
cut-off wavelength scales as $\lambda_{min} \sim 2a$, 
with $a$ being the long side of the rectangular waveguide 
and $b = a/2$ being the short side of the waveguide. 
It is then a straightforward microwave engineering task 
to excite a rectangular waveguide with a Josephson device 
by means of a planar antenna.\cite{westig2011} 
The latter antenna design is compatible with co-planar 
waveguides frequently used in circuit quantum-electrodynamics. 
A second way to excite the fundamental waveguide mode 
in an even simpler way, but only working for lower GHz 
frequencies to about 60~GHz, uses a co-axial cable fixed in 
a back-short waveguide piece along the $E$-plane. For such 
a transition to work with low return loss, the co-axial 
cable inner core is fixed at a distance $\sim b$ away from 
the back-short and extending a distance $\sim b/2$ 
into the waveguide. The latter co-axial cable to waveguide transition 
is a standard and commercially available microwave element.

Connecting the aforementioned Josephson photonic 
devices in this way to the waveguide field, one takes 
advantage of the flexible way of generating tailored 
photonic fields in Josephson devices through strong 
charge-light coupling. At the same time 
one could radiate these fields into free space by means of the 
diagonal-horn antenna where quantum optic 
techniques provide to manipulate and measure the fields in 
a more flexible way than in a circuit configuration. For instance, 
for a free-space field
one can then exploit additionally the polarization degree 
of freedom as reviewed in more detail by \citet{sanz2018_A, sanz2018_B}. 
Free-space ultra low-loss coupling to additional quantum devices, 
for instance to build a network, shows another benefit of this 
scheme.\cite{sanz2018_A, sanz2018_B} In order to minimize the signal loss 
in the immediate vicinity of the Josephson device, important 
to realize a number-state channel or to preserve a high degree of 
amplitude squeezing contained in a two-mode field, the 
rectangular waveguide should be made of a 
superconducting material or has to be 
coated with a superconductor up to a thickness much 
larger than the electro-magnetic penetration depth.
\section{\label{sec:07}Conclusion}
To conclude, we have measured the transmission 
of different length rectangular full-height waveguides 
between 160~GHz and 300~GHz, connected to 
diagonal-horn antennas and optically coupled to 
a coherent detection scheme using a high-impedance 
source and detector with high dynamic range. A detailed 
but still simple enough electro-magnetic model includes the 
optical properties of the diagonal-horn antennas, the dispersion 
relations of the waveguides and the optical properties 
of the source and detector together with the coupling to 
those. The model accounts for most of the measured 
features. The central outcome of our measurements are 
highly-resolved transmission functions which 
reflect the frequency dependent impedance of the 
diagonal-horn and waveguide assembly, and the coupling mismatch to 
the source and the detector. A careful estimation of the 
channel capacity obtains a rate of 1 bit per photon in 
the wideband channel limit of a coherent state. 
\begin{acknowledgements}
We acknowledge funding through the European 
Research Council Advanced Grant No.~339306 
(METIQUM). We like to thank Michael Schultz and the 
precision-machining workshop at the 
I.~Physikalisches Institut of the Universit{\"a}t zu K{\"o}ln 
for expert assistance in the design and the fabrication 
of the diagonal-horn antennas and the waveguides. 
We also thank Anselm Deninger from 
TOPTICA Photonics AG, Germany, for extensive technical 
discussions. MW acknowledges the stimulating support 
from Katharina Franke and her group at the Freie 
Universit{\"a}t Berlin during the completion of this manuscript.
\end{acknowledgements}
\bibliography{paper}
\end{document}